# The Role of AI, Blockchain, Cloud, and Data (ABCD) in Enhancing Learning Assessments of College Students

**RODRIGUEZ, Joel Mark P.[1], AUSTRIA, Genesis S.[2] and MILLAR, Glen B.[3]**

(1) 0009-0009-7195-8164; Rizal Technological University, Mandaluyong City, Philippines, jmprodriguez@rtu.edu.ph
(2) Technological University of the Philippines, Manila City, Philippines
(3) FAITH Colleges: First Asia Institute of Technology and Humanities, Batangas City, Philippines

**Abstract**

This study investigates how ABCD technologies can improve learning assessments in higher education. The objective is to research how students perceive things, plan their behavior, and how ABCD technologies affect individual learning, academic integrity, co-learning, and trust in the assessment. Through a quantitative research design, survey responses were gathered from university students, and statistical tests, such as correlation and regression, were used to establish relationships between Perceived Usefulness (PU), Perceived Ease of Use (PEU), and Behavioral Intention (BI) towards ABCD adoption. The results showed that there was no significant relationship between PU, PEU, and BI, which suggests that students' attitudes, institutional policies, faculty support, and infrastructure matter more in adoption than institutional policies, faculty support, and infrastructure. While students recognize ABCD's efficiency and security benefits, fairness, ease of use, and engagement issues limit their adoption of these technologies. The research adds to Technology Acceptance Model (TAM) and Constructivist Learning Theory (CLT) by emphasizing external drivers of technology adoption. The limitations are based on self-reported data and one institutional sample. It is suggested that universities invest in faculty development, infrastructure, and policy-making to facilitate effective and ethical use of ABCD technologies in higher education.

# INTRODUCTION

The accelerated development of Artificial Intelligence (AI), Blockchain, Cloud, and Data (ABCD) technologies has drastically revolutionized several industries, among them education. AI-based learning tests now give individualized feedback, real-time tracking of performance, and predictive analytics for improving the learning experience of students (Santos & Junior, 2024). Universities are now using these technologies to remain competitive by applying AI, data analytics, and blockchain in the delivery and assessment of courses so that a holistic approach is achieved in the development of students (Williams, 2018). There is still a knowledge gap regarding the potential of ABCD technologies to change the assessment approaches for college students.

While traditional testing has continued to be the norm in gauging scholastic accomplishment, it usually cannot give a fair picture of students' critical and creative thinking capabilities. AI provides an opportunity to bridge the gap by designing real, interactive, and reflective tests that adjust to the demands of modern education (Miserandino, 2024). Nonetheless, most institutions remain in need of digital infrastructure and policies required to support full implementation of ABCD technologies, with issues around privacy of data, ethics, and investment in staff development (Rodriguez, 2024).

This research is especially relevant in the current education environment where technology-enhanced learning is becoming increasingly common. Based on an investigation into the contribution of AI, Blockchain, Cloud, and Data in improving learning tests, this research seeks to identify how these technologies enhance the validity, efficacy, and motivation of tests. It will also assist in arguing that there is a need for robust digital infrastructure, public-private partnerships, and legislation facilitating the uptake of ABCD technology sustainably in higher education.

**Literature Review**

**Overview of Learning Assessments in Higher Education**

Higher education learning examinations have traditionally applied standardized testing devices, and instructors' preferences depend mainly on how educated they are (Saher et al., 2022). Tests of this sort are a vital determinant of other-formation as well as self-formation and play a highly significant role in students' life experiences and the future of their studies (Nieminen & Yang, 2023). Nonetheless, conventional closed-book examinations pose challenges concerning academic integrity, study practices, student stress, performance, and long-term retention of knowledge (Parker et al., 2021). With the development of higher education, advances in technology for learning assessments bring new aspects that increase reliability and validity, and conventional approaches become less time-consuming and less limited (Bahar, 2023). Online assessment tools like Quizziz, Google Forms, and Kahoot have proven indispensable in contemporary education, automating processes of learning, administration, and assessment while facilitating more interactive and effective assessment (Sobirin et al., 2023).

**The Role of Artificial Intelligence and Blockchain for Learning, Secure and Transparent Assessment**

Artificial Intelligence (AI) is transforming learning assessments by enabling subject-matter experts to efficiently process large amounts of data, yielding process profiles that provide a holistic view of students' testing behaviors and overall performance (Guo et al., 2024). Educational assessment technologies powered by AI enhance accuracy, efficiency, and tailored feedback, as well as address challenges and mitigate risks associated with traditional evaluation methods (Owan et al., 2023). In education using computers, AI supports the evaluation of students through differentiated learning opportunities, ongoing measurements, targeted interventions, and generalized gains in overall learning (Santos & Junior, 2023). Blockchain technology, on its own, enhances security, transparency, and confidence for educational evaluations through improved control of data on access, identification verification, accountability, and cost-effective student record management (Razzaq, 2023). It provides a secure and flexible data delivery system, which eliminates trust issues in e-learning and secures sensitive scholarly data (Sastry & Banik, 2021). While blockchain integration into e-learning increases efficiency and transparency, it is not without some of its own constraints and hurdles that must be addressed for wider adoption (Bidry et al., 2023).

**The Cloud Computing and Data Analytics in Learning Assessments and Outcomes**

Cloud computing is central to contemporary learning tests through supporting ongoing assessment of student performance through collaborative cloud-enabled tools, supporting an interactive and data-driven learning environment (Rickards & Steele, 2020). Blending deep learning and cloud computing has been proved to enhance the effectiveness of teaching, raising the interest and initiative of students to learn by 30% and 20%, respectively (Jiang & Sun, 2022).

Partial deployment of cloud-based Knowledge Management Systems also facilitates better quality learning, where students excel above average compared to traditional assessment methods (Liu et al., 2020). Moreover, learning analytics and instructional design have a significant role in enhancing the performance of students through enhanced socio-collaborative and self-learning abilities (Blumenstein, 2020). Teaching Analytics (TA) also improves the quality of instruction by giving teachers actionable data and evidence, walking them through a Teaching Outcome Model (TOM) for more effective engagement with learning metrics (Ndukwe & Daniel, 2020). Moreover, learning analytics enables teachers to evaluate students' conceptions in real-time, tailoring their instructional approaches to offer more focused support and efficient teaching methods (Stanja et al., 2023).

**Challenges and Considerations in Implementing AI, Blockchain, Cloud and Data (ABCD) Technologies in Higher Education**

The integration of AI, Blockchain, Cloud, and Data (ABCD) technologies into higher education comes with various challenges, including those related to privacy, data protection, and ethical issues (Chaka, 2023; Rodriguez, 2024). The incorporation of AI into curriculum programs is a major challenge since institutions have yet to harmonize technological development with conventional pedagogical approaches (Leffia et al., 2024). While AI is speeding up learning through adaptive instruction, interactive engagement, and inclusive learning environments, it generates fear of appropriate analysis of data, algorithmic biases, and appropriate use (Kuleto et al., 2021; Samman, 2024). Similarly, Big Data also has a lot to offer in solving the educational issues but gets its optimal use hindered by privacy concerns, data governance, and organizational readiness (Daniel, 2015). In addition, the success of AI implementation in higher education will not just hinge on whether or not it is capable of improving learning and streamlining administrative procedures but also whether or not it has been appropriately enabled through AI-literate faculty members and morally guided deployment (Murdan & Halkhoree, 2024). In an effort to rise above such threats, investments into digital infrastructure, policy guidelines, and training will be needed so that higher education institutions are made capable and compliant in effectively using ABCD technologies.

**Future Trends and Research Gaps**

The future of AI in tertiary education is characterized by upcoming trends that include hologram technology, ubiquitous learning, automated assessment, green computing, and blended learning pedagogies, all intended to promote educational access and efficiency (Gawande et al., 2020). Blockchain technology is also rapidly being investigated for its uses in academic records management, credential validation, and secure data exchange. Yet, additional research is necessary to combine blockchain with AI, digital innovation, digital maturity, and customer experience to maximize its potential in higher education (Reis-Marques et al., 2021). In spite of the encouraging developments in ABCD technologies, there are gaps in knowing their long-term influence on student learning, institutional take-up, and ethical deployment. Future research needs to be aimed at bridging these gaps through investigating inter-disciplinary models that integrate AI, blockchain, and digital transformation models to achieve a more agile and effective higher education system.

**Theoretical Framework**

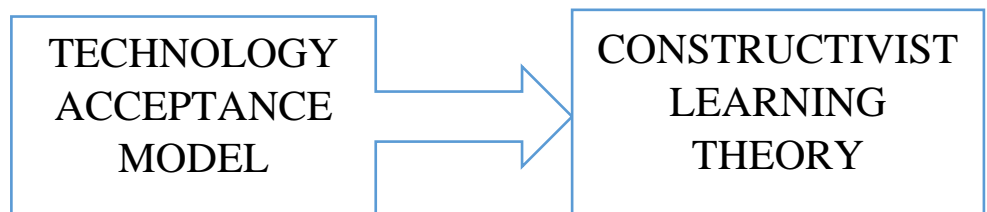

**Figure 1.** *Technology Acceptance Model and Constructivist Learning Theory*

The Technology Acceptance Model (TAM) and Constructivist Learning Theory collectively form a robust theoretical framework for the adoption and efficacy of AI, Blockchain, Cloud, and Data (ABCD) technologies in learning tests. TAM describes how learners perceive, adopt, and plan to use these technologies as a function of their perceived usefulness and ease of use, which has a direct bearing on their willingness to use digital learning tools. At the same time, Constructivist Learning Theory advocates active, student-centered learning where ABCD technologies enhance personalized evaluation, collaborative learning, and knowledge construction. Through the integration of these theories, the research examines not only how students adopt ABCD technologies but also how the technologies improve learning achievements and quality of assessment at university.

**Conceptual Framework**

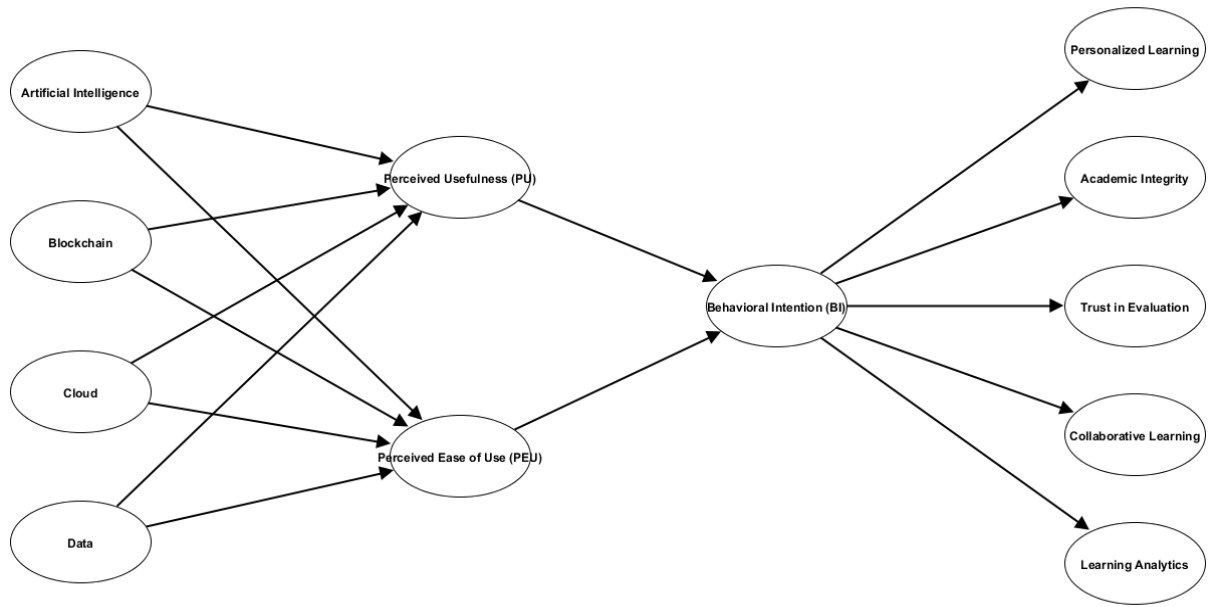

**Figure 2.** *ABCD Technology to Technology Acceptance Model and Constructivist Learning Theory (TAMCOLET) (Rodriguez et al., 2025)*

**Research Question**

1. What is the perception of students toward ABCD technologies in learning assessments in terms of:

    1.1. Artificial Intelligence (AI)

    1.2. Blockchain

    1.3. Cloud Computing

    1.4. Data Analytics

2. How do students perceive AI, Blockchain, Cloud, and Data in terms of their intention to adopt technology-driven learning assessments based on:

    2.1. Perceived Usefulness (PU)

    2.2. Perceived Ease of Use (PEU)

    2.3. Behavioral Intention (BI)

3. How do AI, Blockchain, Cloud, and Data-driven formative assessments support student-centered learning and knowledge construction in higher education in terms of:

    3.1. Personalized Learning

    3.2. Academic Integrity

    3.3. Trust in Evaluation

    3.4. Collaborative Learning

3.5. Learning Analytics

4. Is there a significant relationship between Perceived Usefulness (PU), Perceived Ease of Use (PEOU), and Behavioral Intention (BI) in adopting AI, Blockchain, Cloud, and Data for learning assessments?

5. Does Behavioral Intention significantly predict students' engagement in Personalized Learning, Academic Integrity, Trust in Evaluation, Collaborative Learning, and Learning Analytics?

**Hypothesis**

**HO1:** There is no significant relationship between Perceived Usefulness (PU), Perceived Ease of Use (PEU), and Behavioral Intention (BI) in adopting AI, Blockchain, Cloud, and Data for learning assessments.

**HO2:** Behavioral Intention does not significantly predict students' engagement in Personalized Learning, Academic Integrity, Trust in Evaluation, Collaborative Learning, and Learning Analytics.

## METHODOLOGY

The present research will utilize a causal-comparative research design to analyze the role of ABCD technologies in improving learning assessments among college students. The research will apply quantitative approaches, i.e., survey questionnaires, to gather data on the perceived usefulness, ease of use, and behavioral intention (TAM) of students and their personalized learning, academic integrity, trust in evaluation, collaborative learning, and learning analytics (Constructivist Learning Theory). A standardized questionnaire using established scales will be administered to a random sample of university students with differing exposure to ABCD technologies in their educational exams. Inferential tests such as correlation and regression analysis will be employed in testing variable relationship. It is an appropriate approach because it aligns with TAM's focus on technology adoption and that of Constructivist Learning Theory on learning outcomes, and hence facilitates the analysis of students' adoption of ABCD technologies and how the tools influence their assessment experiences holistically.

**Population and Sampling**

The population sample of this research are the university students from a selected university within the National Capital Region (NCR) in the Philippines, who are taking technology-based learning tests. There will be a stratified sampling approach used to give fair representation by different academic disciplines and year levels because students may have varied exposures and adaptations to AI, Blockchain, Cloud, and Data (ABCD) technologies in their tests. Cochran's formula will be employed to calculate sample size to meet the needs for proper representation of the students while ensuring statistical integrity with 383 identified respondents. Structured survey questionnaires will be the means through which data collection occurs, addressing issues of students' perceived usefulness and ease of use, TAM—behavioral intention, and following personalized learning, academic honesty, trust in the evaluation, teamwork learning, and learning analytics (Constructivist Learning Theory). The application of a stratified sampling strategy is suitable since it will result in an even spread of the participants drawn from various academic levels, thereby ensuring that analysis across ABCD technologies' impact on learning tests is generalizable and balanced.

**Data Gathering Procedure**

The data collection process will begin with acquiring approval and ethical clearance from the selected university of NCR so that ethical standards and data protection policy are adhered to. Pilot testing will be done to ascertain the comprehensibility and believability of the survey questionnaire before mass deployment. Using a stratified sampling technique, members from different fields of study and year levels shall be selected to ensure exposure to AI, Blockchain, Cloud, and Data (ABCD) technologies in testing that is representative. Online and paper surveys shall be distributed, and participants shall be requested to provide informed consent prior to participation. Once data collection is done, the responses will be cleaned and validated to eliminate inconsistencies prior to performing statistical analysis, such as correlation and regression analysis, to investigate relationships among variables. This ensures accuracy, reliability, and thorough analysis of ABCD technologies' engagement in learning exams.

**RESULTS**

*Table 1. Demographic Profile of the Respondents*

| Year Level | Counts | % of Total |
|---|---|---|
| 1st Year | 100 | 26.1 % |
| 2nd Year | 92 | 24.0 % |
| 3rd Year | 94 | 24.5 % |
| 4th Year or higher | 97 | 25.3 % |
| *Course* | *Counts* | *% of Total* |
| Business | 67 | 17.5 % |
| Education | 59 | 15.4 % |
| Engineering | 66 | 17.2 % |
| Health Sciences | 71 | 18.5 % |
| Information Technology | 66 | 17.2 % |
| Others | 54 | 14.1 % |
| *Have Used ABCD Technologies in Learning Assessments* | *Counts* | *% of Total* |
| No | 201 | 52.5 % |
| Yes | 182 | 47.5 % |
| *Frequency of Using Technology-Driven Assessments* | *Counts* | *% of Total* |
| Always | 99 | 25.8 % |
| Often | 83 | 21.7 % |
| Rarely | 99 | 25.8 % |
| Sometimes | 102 | 26.6 % |

As table 1 shows the respondents' demographic makeup suggests quite even distribution by year levels, with the most (26.1%) in the first year and the least (24.0%) in the second year. Respondents' courses are also evenly covered, with the most (18.5%) being recorded by Health Sciences and the least (15.4%) by Education, while the business, Engineering, and Information Technology have similar percentages. On whether ABCD technologies have been applied in the process of learning assessments, the people responding are close to evenly divided, with 52.5% responding they have never applied them, whereas 47.5% confirmed they had applied them. It reflects nearly balanced exposure of these technologies among the students. Regarding the frequency of use of technology-based assessments, responses are also very varied, with "Sometimes" being the most frequent at 26.6%, then "Always" and "Rarely" each at 25.8%, and "Often" at 21.7%. This means that while technology-based assessments are integrated into the learning process, their use among students is not consistent.

*Table 2. Perception of Students towards ABCD Technologies in Learning Assessment*

| *Indicators* | *Mean* | *SD* | *Verbal Interpretation* |
|---|---|---|---|
| *Artificial Intelligence* | | | |
| 1. AI-based assessments provide real-time feedback that helps improve my learning. | 2.49 | 1.12 | Disagree |
| 2. AI enhances personalized learning experiences by adapting to my strengths and weaknesses. | 2.45 | 1.11 | Disagree |
| 3. AI-driven assessment tools (e.g., automated grading, chatbots) improve efficiency. | 2.56 | 1.14 | Agree |
| 4. AI reduces bias in assessments by providing objective evaluation. | 2.54 | 1.17 | Agree |
| 5. I trust AI-assisted grading to be fair and accurate. | 2.48 | 1.13 | Disagree |
| **OVERALL MEAN** | **2.50** | **1.13** | **Agree** |
| *Blockchain* | | | |
| 1. Blockchain ensures secure storage and verification of academic records. | 2.50 | 1.16 | Agree |
| 2. Blockchain helps maintain assessment integrity by preventing data manipulation. | 2.66 | 1.10 | Agree |
| 3. Using blockchain in learning assessments increases trust and transparency. | 2.54 | 1.11 | Agree |
| 4. I am confident that blockchain-based credentials are tamper-proof and credible. | 2.50 | 1.14 | Agree |
| 5. Blockchain improves fairness in grading and certification processes. | 2.52 | 1.14 | Agree |
| **OVERALL MEAN** | **2.54** | **1.13** | **Agree** |
| *Cloud* | | | |
| 1. Cloud-based platforms (e.g., Google Classroom, Moodle) make assessments more accessible. | 2.45 | 1.14 | Disagree |
| 2. Cloud technology allows me to collaborate effectively with peers in assessments. | 2.56 | 1.14 | Agree |
| 3. I find cloud-based assessments more flexible and convenient than traditional methods. | 2.44 | 1.08 | Disagree |
| 4. Cloud platforms enhance data security and backup for my assessments. | 2.54 | 1.12 | Agree |
| 5. The use of cloud technology in assessments improves my learning experience. | 2.54 | 1.09 | Disagree |
| **OVERALL MEAN** | **2.50** | **1.11** | **Agree** |
| *Data* | | | |
| 1. Learning analytics help me track my academic performance over time. | 2.47 | 1.13 | Disagree |
| 2. Data-driven assessments provide personalized recommendations for my improvement. | 2.55 | 1.13 | Agree |
| 3. Data analytics help instructors provide targeted feedback based on student performance. | 2.50 | 1.12 | Agree |
| 4. I believe data analytics improve assessment accuracy and fairness. | 2.46 | 1.09 | Disagree |
| 5. The integration of learning analytics in assessments enhances my motivation to learn. | 2.61 | 1.12 | Agree |
| **OVERALL MEAN** | | | |

*Legend: "1.00-1.75 Strongly Disagree", "1.76-2.50 Disagree", "2.51-3.25 Agree", "3.26-4.00 Strongly Agree*

The attitude of students toward the ABCD (Artificial Intelligence, Blockchain, Cloud, and Data) technologies in assessment learning is overall positive as measured by means of agreement in all four categories. Students disagreed that Artificial Intelligence would give real-time feedback (2.49), make learning personal (2.45), and prevent bias in grading (2.48), but were in agreement that it would make efforts more efficient (2.56) and minimize bias (2.54). This reflects skepticism about the fairness of AI-based tests but recognition of their effectiveness. In the Blockchain category, all the markers were scored above 2.51, reflecting agreement that blockchain enhances assessment integrity (2.66), increases trust (2.54), and enhances fairness (2.52), which shows that students perceive blockchain as an effective tool for academic record security and authentication.

The Cloud category was inconsistent in its opinions with students splitting their opinions regarding convenience (2.44) and enhancing learning experience (2.50) for cloud-based exams but unconditionally agreeing that cloud facilitates collaboration (2.56), security (2.54), and accessibility (2.45). It is an indicator of having enjoyed the cloud-based assessment system but protested its convenience and effect on learning. Lastly, in the Data Analytics area, students conflicted that learning analytics assist in monitoring academic performance (2.47) and increasing accuracy and equity (2.46) but concurred on data-driven recommendations (2.55) and encouragement (2.61). It means that even though students acknowledge some use of data analytics for customized recommendations, they are still skeptical about its effect on performance monitoring and equity. In general, students have a positive attitude towards ABCD technologies, but fairness, personalization, and flexibility issues still exist.

*Table 3. Students Intention to Adopt ABCD Technologies in Learning Assessment*

| Indicators | Mean | SD | Verbal Interpretation |
|---|---|---|---|
| **Perceived Usefulness (PU)** | | | |
| 1. AI, Blockchain, Cloud, and Data make learning assessments more efficient. | 2.52 | 1.09 | Agree |
| 2. Using these technologies reduces human error in grading and evaluation. | 2.48 | 1.03 | Disagree |
| 3. The integration of ABCD in assessments enhances student engagement. | 2.49 | 1.1 | Disagree |
| 4. These technologies help me understand learning concepts better. | 2.50 | 1.11 | Agree |
| 5. I find ABCD technologies beneficial for improving academic performance. | 2.49 | 1.12 | Disagree |
| **OVERALL MEAN** | **2.50** | **1.09** | **Agree** |
| **Perceived Ease of Use (PEOU)** | | | |
| 1. AI-based assessments are easy to navigate and use. | 2.47 | 1.12 | Disagree |
| 2. Blockchain-secured assessments are simple to understand and access. | 2.46 | 1.12 | Disagree |
| 3. Cloud-based assessments require minimal effort to complete. | 2.45 | 1.11 | Disagree |
| 4. Learning analytics dashboards provide clear and meaningful insights. | 2.57 | 1.11 | Agree |
| 5. I find ABCD-based assessment tools easy to use in my academic activities. | 2.45 | 1.15 | Disagree |
| **OVERALL MEAN** | **2.48** | **1.12** | **Disagree** |
| **Behavioral Intention (BI)** | | | |
| 1. I am open to using AI, Blockchain, Cloud, and Data in my learning assessments. | 2.49 | 1.12 | Disagree |
| 2. I would recommend the use of ABCD technologies for learning assessments. | 2.50 | 1.11 | Agree |
| 3. If given a choice, I prefer ABCD-powered assessments over traditional methods. | 2.55 | 1.12 | Agree |
| 4. My university should increase its use of ABCD technologies in assessments. | 2.46 | 1.11 | Disagree |
| 5. I feel motivated when using technology-based assessments. | 2.45 | 1.13 | Disagree |
| **OVERALL MEAN** | **2.49** | **1.12** | **Disagree** |

*Legend:* "1.00-1.75 Strongly Disagree", "1.76-2.50 Disagree", "2.51-3.25 Agree", "3.26-4.00 Strongly Agree

The examination of students' attitudes to embracing ABCD technologies for learning evaluations shows mixed views, with overall disagreement across the majority of categories.

Regarding Perceived Usefulness (PU), students concur that ABCD technologies enhance efficiency (2.52) and facilitate concept understanding (2.50), but they disagree over whether these technologies minimize human error (2.48), improve engagement (2.49), or positively impact academic performance (2.49). This suggests that while students mention some advantages of ABCD technologies, they are still skeptical of their general effects on their school life. Students firmly disagree on AI-based test ease of use (2.47), ease of blockchain tests (2.46), ease of cloud tests (2.45), and ease of application of ABCD-based tools to academic tasks (2.45). Nevertheless, they concur that learning analytics dashboards yield valuable insights (2.57), reflecting the preference for tools with clear presentation of data. In Behavioral Intention (BI), students mostly disagree with items about their willingness to use AI, Blockchain, Cloud, and Data (2.49), suggesting its application (2.50), and more institutional use (2.46). This implies that although students cite some benefits of ABCD technologies, they remain doubtful of their overall impact on their life in school. Students strongly disagree with AI-based ease of test use (2.47), ease of blockchain tests (2.46), ease of cloud tests (2.45), and ease of use of ABCD-based tools for academic purposes (2.45).

*Table 4. The Role of AI, Blockchain, Cloud, and Data in Student-Centered Learning & Knowledge Construction*

| Indicators | Mean | SD | Verbal Interpretation |
|---|---|---|---|
| **Personalized Learning** | | | |
| 1. AI-driven assessments adjust to my learning pace and style. | 2.52 | 1.14 | Agree |
| 2. Personalized feedback from ABCD-based assessments helps me improve my performance. | 2.57 | 1.14 | Agree |
| 3. These technologies make assessments interactive and engaging. | 2.48 | 1.14 | Disagree |
| 4. ABCD-powered learning tools help me retain knowledge better. | 2.48 | 1.12 | Disagree |
| 5. I feel more engaged when using personalized learning technologies. | 2.49 | 1.13 | Disagree |
| **OVERALL MEAN** | **2.51** | **1.13** | **Agree** |
| *Academic Integrity* | | | |
| 1. Blockchain-based assessments prevent academic dishonesty and fraud. | 2.62 | 1.12 | Agree |
| 2. AI-driven proctoring tools help reduce cheating in online assessments. | 2.42 | 1.11 | Disagree |
| 3. Secure data management in ABCD assessments ensures fairness in grading. | 2.45 | 1.14 | Disagree |
| 4. The use of ABCD technologies increases trust in academic results. | 2.48 | 1.14 | Disagree |
| 5. I believe academic institutions should adopt more secure digital assessment tools. | 2.50 | 1.17 | Agree |
| **OVERALL MEAN** | **2.49** | **1.14** | **Disagree** |
| **Trust in Evaluation** | | | |
| 1. AI-assisted grading is transparent and reliable. | 2.50 | 1.15 | Agree |
| 2. Blockchain technology ensures my grades and academic records are credible. | 2.54 | 1.10 | Agree |
| 3. Cloud-based assessment tools keep my academic data safe. | 2.48 | 1.13 | Disagree |
| 4. Data-driven assessments help ensure fair evaluation of student performance. | 2.52 | 1.13 | Agree |
| 5. The use of ABCD-powered assessments improves my confidence in grading systems. | 2.56 | 1.10 | Agree |
| **OVERALL MEAN** | **2.52** | **1.12** | **Agree** |
| *Collaborative Learning* | | | |
| 1. Cloud-based assessments enhance teamwork and knowledge-sharing. | 2.50 | 1.15 | Agree |
| 2. AI-powered discussion tools support collaborative learning with my peers. | 2.56 | 1.06 | Agree |
| 3. Blockchain enables peer-reviewed assessments that are fair and reliable. | 2.53 | 1.15 | Agree |
| 4. Digital assessments help me develop critical thinking and problem-solving skills. | 2.52 | 1.14 | Agree |
| 5. Group assessments using ABCD tools improve my ability to work in teams. | 2.51 | 1.13 | Agree |
| **OVERALL MEAN** | **2.49** | **1.12** | **Disagree** |
| *Learning Analytics* | | | |
| 1. AI-driven learning analytics help me track my academic performance. | 2.49 | 1.08 | Disagree |
| 2. Data analytics tools provide recommendations to improve my learning strategies. | 2.43 | 1.08 | Disagree |
| 3. AI-generated reports help me adjust my study habits effectively. | 2.48 | 1.12 | Disagree |
| 4. Cloud-based learning analytics allow instructors to provide better feedback. | 2.51 | 1.12 | Agree |
| 5. Using learning analytics boosts my motivation to improve academically. | 2.47 | 1.12 | Disagree |
| **OVERALL MEAN** | **2.48** | **1.12** | **Disagree** |

Analysis of AI, Blockchain, Cloud, and Data's role in student-led learning and construction of knowledge comes across with an overall mixed sentiment among the students, wherein certain elements found endorsement while others have met criticism. Under Personalized Learning, the students were on board that assessments using AI adjust to the learners' speed (2.52) and personalized feedback supports improvement in performance (2.57). However, they differed regarding whether such technologies improve interactivity (2.48), contribute to knowledge retention (2.48), or improve engagement (2.49). This implies that although they appreciate the worth of individualized feedback, students are not sure about its net effect on motivation and retention.

In Academic Integrity, the respondents were convinced that blockchain exams ensure no cheating (2.62) and ABCD ensures trust in results (2.48), but not so confident whether AI-based monitoring can avoid cheating (2.42) or ABCD ensures fair grading (2.45). It indicates higher trust in blockchain in offering assurance of integrity than for AI-based monitoring. For Trust in Evaluation, the students in general agreed that AI grading is transparent (2.50), blockchain adds credibility to records (2.54), and cloud storage preserves data security (2.48). They were also in agreement that data-driven judgment provides equitable judgments (2.52) and adds to the trust in grading processes (2.56), which is in general an indicator of trusting the role of ABCD in providing credible assessment processes. In Collaborative Learning, the students appreciated AI-assisted discussion tools to group work (2.56) and cloud tests to sharing information (2.50). They also concurred on the fairness of peer-reviewed tests (2.53) and computer-based tests to critical thinking (2.56), even though they were not so sure if ABCD tools encourage teamwork skill (2.49). Lastly, under Learning Analytics, students were largely skeptical whether it can monitor academic performance (2.49) or provide advice (2.43), even though they admitted study habits are assisted through reports submitted by AI (2.52). Overall, even while students envision promise in ABCD technologies regarding integrity of testing, trust in grades, and collaboration, skepticism remains about their effectiveness in engagement, fairness, and learning analytics.

*Table 5. The Role of AI, Blockchain, Cloud, and Data in Student-Centered Learning & Knowledge Construction*

| | | Pearson's r | P value | Decision on Ho | Interpretation |
|---|---|---|---|---|---|
| Perceived Usefulness | - Perceived Ease of Use | -0.006 | 0.911 | Accepted | There is no significant relationship between Perceived Usefulness (PU), Perceived Ease of Use (PEU), and Behavioral Intention (BI) in adopting AI, Blockchain, Cloud, and Data for learning |
| Perceived Usefulness | - Behavioral Intention | 0.035 | 0.495 | | |
| Perceived Ease of Use | - Behavioral Intention | 0.022 | 0.667 | | |

The results shown in Table 5 show there is no existence of a relationship that is statistically significant between Behavioral Intention (BI), Perceived Ease of Use (PEU), Perceived Usefulness (PU) and adoption of AI, Blockchain, Cloud, and Data in learning. Pearson's r values are extremely close to zero (-0.006, 0.035, and 0.022), while the p-values (0.911, 0.495, and 0.667) are significantly higher than the universal significance level of 0.05, and hence the null hypothesis (Ho) is accepted. This indicates that students' beliefs regarding how easy or helpful ABCD technologies are do not have a significant effect on their adoption intention for learning. That is, regardless of the extent to which students view such technologies as helpful or easy to use, they do not always end up applying them in their own learning. This might indicate that the effects of other variables, including institution-based support, computer proficiency, or learning disposition, could dominate to influence behavior concerning adoption. Following this, educational use of ABCD technologies could be encouraged by measures that will have to deal with extrinsic motivators and not just utility and simplicity only.

*Table 6. The Behavioral Intention as Predictors in Student-Centered Learning & Knowledge Construction*

| | Unstandardized | Standard Error | Standardized | t | p | Collinearity Statistics | |
|---|---|---|---|---|---|---|---|
| | | | | | | Tolerance | VIF |
| (Constant) | 2.595 | 0.291 | | 8.93 | < .001 | | |
| Personalized Learning | -0.032 | 0.052 | -0.032 | -0.626 | 0.532 | 0.995 | 1.006 |
| Academic Integrity | 0.064 | 0.053 | 0.063 | 1.219 | 0.224 | 0.997 | 1.003 |
| Trust in Evaluation | -0.009 | 0.051 | -0.009 | -0.173 | 0.862 | 0.994 | 1.006 |
| Collaborative Learning | -0.021 | 0.055 | -0.019 | -0.374 | 0.708 | 0.998 | 1.002 |
| Learning Analytics | -0.044 | 0.054 | -0.042 | -0.822 | 0.412 | 0.995 | 1.005 |

Dependent Variable: ***Behavioral Intention***

R-squared: 0.007; F value: 0.531, p value: 0.753

The regression presented in Table 6 tests the model of if and how Personalized Learning, Academic Integrity, Trust in Assessment, Collaborative Learning, and Learning Analytics influence Behavioral Intention on student-based learning and construction of knowledge. All the predictors yield p-values which are all several orders above the conventional critical level of significance for those predictors an influential effect statistically, i.e., at \alpha = 0.05 or less on each. The value of R-squared (0.007) indicates that only 0.7% variation in behavioral intention is accounted for by these predictors, reflecting a very weak model. Moreover, the F-value (0.531) and p-value (0.753) also validate that the model as a whole is not statistically significant. The collinearity statistics of Tolerance and VIF reflect no multicollinearity issues because all the VIFs are near 1, which means independent variables are not strongly correlated. These results indicate that students' behavioral intention to use AI, Blockchain, Cloud, and Data in learning is not affected by perceived personalized learning, academic integrity, trust in assessment, collaborative learning, or learning analytics.This agrees with previous studies in which perceived ease of use and usefulness also had no influence on behavioral intention. Therefore, extrinsic aspects such as institutional policy, infrastructure for digital use, or prior use of technology by students may play a stronger role in deciding their readiness to adopt ABCD technologies.

## DISCUSSION

The results of the present research support previous research in the complexity of implementing AI, Blockchain, Cloud, and Data (ABCD) technologies for learning assessments in higher education. Although technologies such as Quizziz, Google Forms, and Kahoot have revolutionized testing (Sobirin et al., 2023), learners in this research displayed reluctance to entirely adopt ABCD technologies, as seen in the non-significant associations between Perceived Usefulness (PU), Perceived Ease of Use (PEU), and Behavioral Intention (BI). Blockchain and AI, as promising as they are to heighten security, transparency, and automated grading (Razzaq, 2023; Owan et al., 2023), were not viewed as having an effect on the students' adoption intention. In the same way, cloud computing and data analytics, proven to enhance learning achievement and interest (Rickards & Steele, 2020; Blumenstein, 2020), were not observed to significantly affect behavioral intention, and this might imply that students are not yet fully confident in or aware of their benefits. One of the principal challenges in using ABCD technologies is the disconnect between technological progress and traditional pedagogical paradigms (Leffia et al., 2024), as well as issues surrounding algorithmic biases, data protection, and institutional preparedness (Kuleto et al., 2021; Bidry et al., 2023). In addition, regression analysis indicated that Personalized Learning, Academic Integrity, Trust in Evaluation, Collaborative Learning, and Learning Analytics were not significant predictors of behavioral intention, highlighting the fact that adoption could be contingent on institutional policy, digital infrastructure, and training of faculty instead of students' perceptions. Looking forward, whereas AI-driven tests and blockchain-enabled credentialing will continue to mold higher education (Gawande et al., 2020; Reis-Marques et al., 2021), there are research gaps as to their long-term effects on student engagement and institutional uptake. Investments in digital infrastructure, ethical AI deployment, and faculty training will be required to close the gap between technological developments and effective pedagogy in higher education in order to achieve successful implementation.

**THEORETICAL IMPLICATIONS**

The findings of the study contribute to the theoretical grounding of Technology Acceptance Model (TAM) and Constructivist Learning Theory (CLT) in AI, Blockchain, Cloud, and Data (ABCD) technologies within learning tests. TAM assumes perceived usefulness and perceived ease of use as predictors of behavioral intention; however, such was not obtained in this research where these aspects did not bear a significant link with the inclination of students toward adopting ABCD technologies. This contests current literature with the belief in a direct correlation between perceived usefulness and ease of technology adoption. Rather, behavioural intention could be influenced to a greater extent by external institutional enablers, digital infrastructure, and technology prior exposure. From a constructivist point of view, although ABCD technologies hold the promise of improving personalized learning, collaboration, and assessment integrity, student cynicism indicates that these advantages are not yet fully achieved in present educational practice. The study therefore contributes to theoretical knowledge by proposing that institutional preparedness, pedagogical congruence, and faculty development are mediators required in ABCD adoption as opposed to user perception alone.

**LIMITATIONS**

The findings of this research have several shortcomings that must be considered while drawing conclusions from them. First, the study samples are derived from students of one specific Philippine institution, and it may thereby impose limitations on making generalizations based on the results to other academies. Second, the study was based on self-report data and thus susceptible to such influences as social desirability or lack of awareness of the nuances of ABCD technologies. Third, quantitative survey data were used in this research, and although statistically significant, do not have rich qualitative data on motivations, experiences, or challenges in using ABCD technologies by students. Fourth, the study did not include faculty opinions, institution policy, or technology infrastructure, which might be a more equitable reflection of challenges and opportunities faced because of ABCD implementation. Future studies can make use of mixed-method designs, increase the sample to more than one institution, and utilize the faculty and administrative voices to provide further illumination.

**CONCLUSION**

Accordingly, this research has examined students' perception and willingness to act toward the use of AI, Blockchain, Cloud, and Data technologies in assessment learning. Results suggest that students acknowledge the probable advantage of ABCD technologies in improving efficiency, security, and cooperation but have a skeptical view on their usability, fairness, and general contribution to learning interaction. The absence of a strong relationship between Perceived Usefulness (PU), Perceived Ease of Use (PEU), and Behavioral Intention (BI) indicates that institution-related, pedagogic, and infrastructural elements have a stronger determining influence on students' adoption behaviors than their individual perceptions. In addition, regression analysis indicated that Personalized Learning, Academic Integrity, Trust in Evaluation, Collaborative Learning, and Learning Analytics were not significant predictors of behavioral intention, supporting the idea that factors outside of technology may be more impactful. Thus, effective adoption of ABCD technologies will involve more than having technology available—it requires faculty development, institutional encouragement, and enhanced digital infrastructure.

**RECOMMENDATION**

To enable the use and implementation of AI, Blockchain, Cloud, and Data (ABCD) technologies in education testing, higher education institutions must invest in quality digital infrastructure as well as clearly defined policies on ethics, security, and data privacy. Training programs for faculty members should be instituted to enhance digital literacy and integration of ABCD technologies into the curriculum through pilot projects, experiential learning, and faculty-student teams. At the same time, students need to be exposed more to ABCD technologies through awareness programs, workshops, and practical experiences that emphasize their advantages in personalized learning, academic honesty, and participation. The emphasis should also be on maintaining transparency and fairness when AI-based or blockchain-protected tests are done, through ongoing audits of the AI-based systems of grading for avoiding prejudice and applying blockchain in tamper-resistant, secure, and transparent study records. It is important that institutions weigh models of hybrid assessing that combine human checking with AI-based grading in order to make them fair and reliable. More policy formulation and research are also needed to analyze the long-term effect of ABCD technologies on students' participation and institutional adoption as well as the formulation of uniform guidelines for covering ethics, privacy, accessibility, and infrastructure issues. Through the adoption of such approaches, institutions can fill the gap between technological innovation and pedagogic responsiveness so that ABCD technologies facilitate learning processes in students directly, instead of making more problems.

**Ethics Statement**

The participants in this study have given their informed consent without delay to the data collection, data handling, and data disposal process to be followed in conducting the research.